 \documentclass[prl,floats,showpacs,onecolum,twocolumn]{revtex4-1}

\usepackage[english]{babel}
\usepackage{amsmath}
\usepackage{amssymb, comment}
\usepackage[]{graphicx}
\usepackage{times}
\usepackage{bm}
\usepackage{units}
\usepackage{braket}
\usepackage[version=3]{mhchem}

\usepackage{bbm}

\usepackage{bm}
\usepackage{amsmath}
\usepackage{amssymb, comment}

\renewcommand{\eqref}[1]{Eq. (\ref{#1})}

\usepackage{epstopdf}
\usepackage{array}

\usepackage{color}

\renewcommand{\eqref}[1]{Eq. (\ref{#1})}

\begin{document}
\title{Optical fingerprint of bright and dark localized excitonic states in atomically thin 2D materials }

\author{Maja Feierabend, Samuel Brem and Ermin Malic}
\address{Chalmers University of Technology, Department of 
Physics, 412 96 Gothenburg, Sweden}


\begin{abstract}
Point defects, local strain or impurities can crucially impact the optical response of atomically thin two-dimensional materials as they offer trapping potentials for excitons. These trapped excitons appear in photoluminescence spectra as new resonances below the bright exciton that can even be exploited for single photon emission. While large progress has been made in deterministically introducing defects, only little is known about their impact on the optical fingerprint of 2D materials.
Here, based on a microscopic approach we reveal direct signatures of localized bright excitonic states as well as indirect phonon-assisted side bands of localized momentum-dark excitons. The visibility of localized excitons strongly depends on temperature and disorder potential width. This results in different regimes, where either the bright or dark localized states are dominant in optical spectra. We trace back this behavior to an interplay between disorder-induced exciton capture and intervalley exciton-phonon scattering processes.
\end{abstract}

\maketitle
\section{Introduction}
Monolayer transition metal dichalcogenides (TMDs) are remarkable materials, among others due to their exceptional optical properties \cite{erminnpj,alexeyReview}. Their optical response at room temperature is dominated by optically accessible bright excitons and the resulting resonances in optical spectra are well understood based on strong Coulomb and light-matter interactions \cite{THeinz,gunnar_prb}.  The excitonic resonances were shown to be controllable by strain, chemical functionalization and doping \cite{island2016precise,conley2013bandgap,ross2014electrically,mouri2013tunable}. 
However, at low temperatures the optical response is much more complicated. It is dominated by a variety of resonances energetically below the bright excitons, which have not been fully understood yet. While some of those peaks can be explained by trions \cite{godde2016exciton, jadczak2017probing}, biexcitons \cite{nagler2018zeeman} or momentum- and spin-dark excitons \cite{malic2018dark,molas2017brightening,zhou2017probing,hoegeleMono,zhang2015experimental,brem2019phonon}), little is known about the origin of defect-determined emission. The latter appears at holes \cite{kumar2015strain}, etched surfaces\cite{rosenberger2019quantum}, edges \cite{koperski2015single}, nanopillars \cite{palacios2017large},  areas of local strain \cite{kern2016nanoscale}, nanobubbles \cite{shepard2017nanobubble} or similar. Moreover, defect-related emission resonances have great potential as sources for single-photon emitters \cite{chakraborty2015voltage,koperski2015single,he2015single,srivastava2015optically,tonndorf2015single}
and hence a more fundamental understanding of the origin of these peaks is crucial.

\begin{figure}[t!]
  \begin{center}
\includegraphics[width=\linewidth]{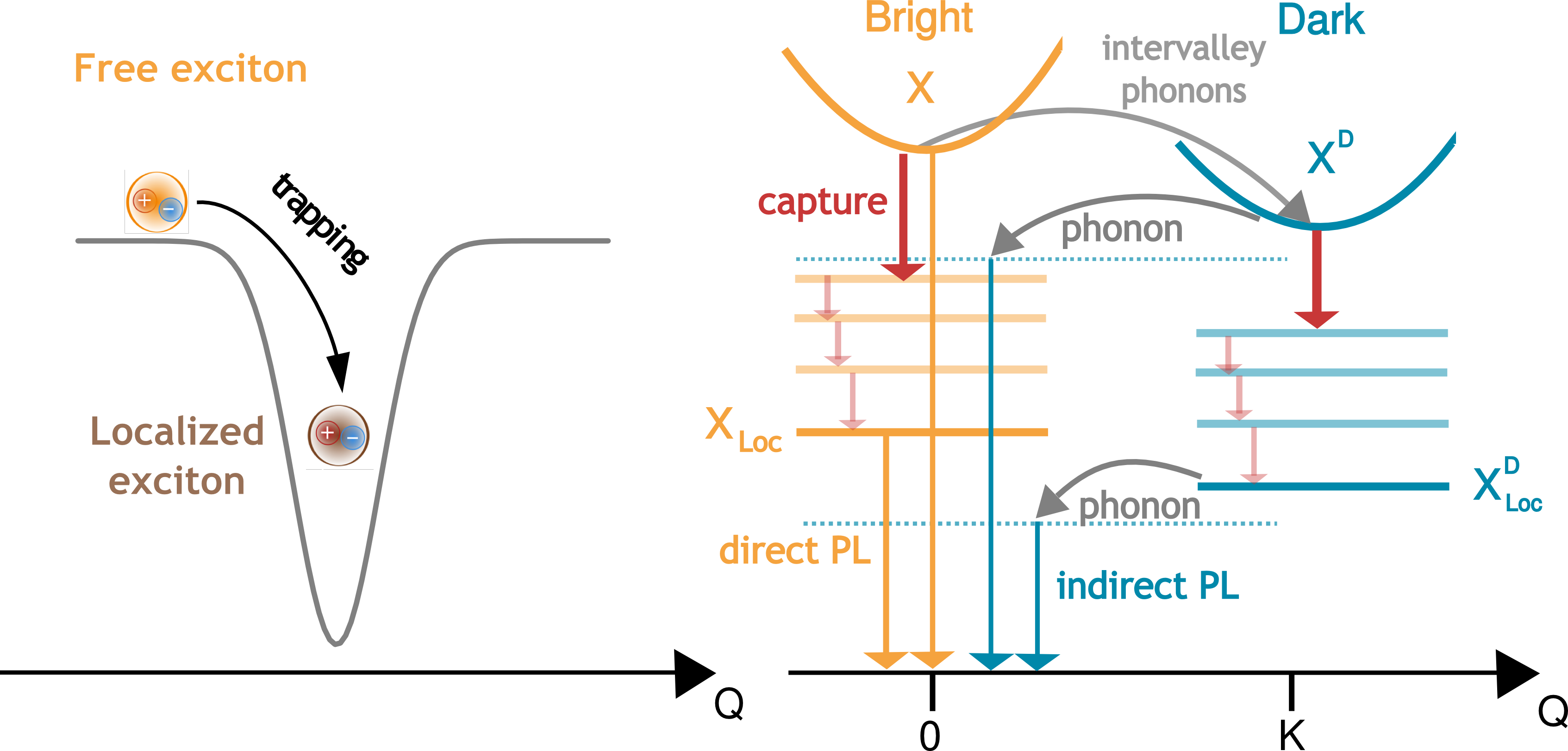} 
\end{center}
    \caption{\textbf{Disorder trapping and photoluminescence of excitons.} (a) Excitons can become trapped in the vicinity of  disorder potentials that are broad and deep enough.  The  binding energy and wavefunction of these localized excitons depend on the characteristics of the disorder potential.  (b) Phonon-driven relaxation and capture processes involving both bright $X$ and momentum-dark $X^D$ excitons as well as their localized states $X_{\text{Loc}}$ and $X^{D}_{\text{Loc}}$, respectively. We include phonon-assisted intervalley scattering, capture and relaxation of excitons as well as direct and phonon-assisted photoluminescence. }
   \label{schema}
\end{figure}

In this work, we shed light on the temperature-dependent optical fingerprint of TMDs spectrally below the bright exciton in presence of defects. We include direct photoluminescence (PL) from the bright exciton $X$ and its localized states $X_{\text{Loc}}$ as well as phonon-assisted PL from momentum-dark excitons $X^D$ and the corresponding localized states $X^D_{\text{Loc}}$, cf. Fig. \ref{schema}. We model both free and localized excitons on the same microscopic footing. While free excitons are formed by a Coulomb potential leading to the quantization of the relative motion of electron and hole, localized excitons  are formed due to trapping into a disorder potential  giving rise to a quantization of the exciton center-of-mass motion. We show that depending on the width of the trapping potential, the existence of localized states as well  as exciton capture and recombination rates can be controlled. The calculated optical spectra agree well with recent experiments, observing localized states due to local strain  \cite{kumar2015strain,kern2016nanoscale} or in presence of nanopillars \cite{branny2017deterministic,palacios2017large}. Moreover,  depending on temperature and disorder potential characteristics we predict phonon- or localization-dominated regimes in PL spectra of TMDs. 

\section{Theoretical approach}
To describe the optical fingerprint of free and localized excitonic states on a microscopic level, we exploit the density matrix formalism  \cite{Kochbuch, Kira2006,carbonbuch}, where we apply the cluster expansion approach in excitonic basis \cite{gunnar_prb,selig2018dark,brem2018exciton}. This allows us to describe the entire exciton landscape including bright and dark exciton states, their binding energies, wavefunctions and signatures in PL spectra. 
The main goal of our study is to investigate under which conditions a disorder potential can trap excitons and how it influences the optical fingerprint of TMDs including signatures of bright and momentum-dark excitonic states. 

To obtain access to the optical response of these materials, the knowledge of the number of emitted photons $ n_q=\langle c^{\dagger}_qc_q\rangle$ is crucial as it determines the steady-state photoluminescence $\text{PL}(\omega_q)=\omega_q \frac{\partial}{\partial t} n_q$. The dynamics of the photon number $n_q$ on the other side depends on the photon-assisted polarization \cite{thranhardt2000quantum,brem2019phonon} 
  $S^{v c}_{\bf{k_1}\bf{k_2}}(t)=\langle c_q^\dagger a_{\bf k_1}^{\dagger v} a_{\bf k_2}^{c} \rangle$ resultign in  $\text{PL}(\omega_q) \propto \frac{\partial}{\partial t} n_q \propto S^{v c}_{\bf{k_1}\bf{k_2}}(t)$. This microscopic quantity is a measure for optically induced transitions from the state $(v_,\bf k_1)$ to the state $(c,\bf k_2) $ under annihilation (creation) of a photon $c_q^{(\dagger)}$. The states are characterized by the electronic momentum $\bf k_i$, and the band index $\lambda=v,c$ denoting valence or conduction band, respectively. Note that we take into account the conduction band minima at the K,$\Lambda,$ and K' valley, which are crucial for the formation of momentum-dark exciton states \cite{malic2018dark}. 

To account for excitonic effects, which are dominant in these materials \cite{Chernikov2014, gunnar_prb,arora2015excitonic,erminnpj}, we project the photon-assisted polarization into an excitonic basis.
 We use the relation
$S_{\bf{qQ}}^{vc }= \sum_{\mu} \varphi_{\bf q}^{\mu} S_{\bf{Q}}^{\mu}$, where our original observable $S_{\bf{qQ}}^{vc }$ is projected to a new excitonic quantity $S_{\bf{Q}}^{\mu}$ that is weighted by the excitonic wave function $\varphi_{\bf q}^{\mu}$.
Here, we have introduced the center-of-mass momentum $\bf Q = k_2 - k_1$ and the relative momentum ${\bf q}=\alpha {\bf k_1} + \beta {\bf k_2}$ with  $\alpha= \frac{m_h}{m_h + m_e^{}}$ and $\beta=\frac{m_e^{}}{m_h + m_e^{}}$  with the electron (hole) mass $m_{e(h)}^{}$.
 The free excitonic eigenfunctions $\varphi_{q}$ and eigenenergies $\varepsilon^{\mu}$ are obtained by solving the Wannier equation, which presents an eigenvalue problem for excitons \cite{Kochbuch, Kira2006,gunnar_prb,brem2019phonon,malic2018dark}. The corresponding Coulomb matrix elements are calculated using a  Keldysh potential \cite{Keldysh1978, gunnar_prb,Cudazzo2011}.
 
To obtain  the temporal evolution of $S_{\bf{Q}}^{\mu}(t)$, we exploit the Heisenberg equation of motion $i\hbar \dot S_{\bf Q}^\mu (t)=[H,S_{\bf Q}^\mu (t)]$ \cite{Kochbuch, carbonbuch}, which requires the knowledge of the many-particle Hamilton operator $H$. The latter reads in this work $H = H_0 + H_{c-l} + H_{c-phon} + H_{c-dis}$ including the free carrier contribution $H_0$, the carrier-light interaction $H_{c-l}$, the carrier-phonon coupling  $H_{c-phon}$ and the carrier-disorder interaction $H_{c-dis}$. To calculate the matrix elements, we apply the nearest-neighbor tight-binding approach \cite{ermin_cm,Kochbuch,Kira2006} including fixed (not adjustable) input parameters from DFT calculations of the electronic bandstructure \cite{andor}.

The crucial part of the Hamilton operator in this work is the carrier-disorder matrix element $g^{c-dis}_{k_1,k_2}=\langle\Psi_{k_1}({\bf r}) | V_{\text{dis}}({ \bf r}) |  \Psi_{k_2}( {\bf r}) \rangle$ with a Gaussian disorder potential $
 V_{\text{dis}}({\bf r}) = V_0 \exp{\left(\frac{-4\ln 2  ( { \bf r}-{\bf R}_0)^2}{\sigma^2}\right)}
$ 
 with $\sigma$  denoting the full-width half maximum in real space,  $V_0$ the depth of the potential and $R_0$ the position of the disorder. Assuming disorder centers far away from each other, $R_0$ only leads to a phase in the momentum space. 
The approximation of a Gaussian disorder potential is well established in theory \cite{wen2005binding,hichri2017exciton,adamowski2000electron} and moreover, experimentally measured potential traps reveal a Gaussian-like behavior \cite{rosenberger2019quantum,kern2016nanoscale}.
Note that our theory is not limited to  Gaussian potentials but could be extended to other potential forms, such as elliptical \cite{rosati2018spatial} or nanobubbles \cite{brooks2018theory}. The investigated Gaussian potential in its general form can have its origin in a variety of physical phenomena, such as local strain gradients stemming e.g. from nanopillars on TMD monolayers \cite{branny2017deterministic,kern2016nanoscale,palacios2017large} but also in defect- or impurity-induced potentials on  atomic level \cite{tran2017room}. A discussion on certain regimes and comparison with experimental observations follows in the results section.

By transforming the disorder potential into momentum space we find that it leads to a quantization of the exciton center-of-mass momentum, i.e. excitons cannot move freely anymore \cite{hichri2017exciton}. Then, we project the phonon-assisted polarization into a  localized excitonic basis with $S_{\bf{Q}}^{\mu}(t) =\sum_ {n} \chi^{\mu n}_{\bf Q}S^{\mu n}(t)$, where $\chi^{\mu n}_{\bf Q}$ are the localized excitonic wavefunctions. Note that all types of free excitons including bright KK or momentum-dark intervalley K$\Lambda$ and KK' excitons \cite{berghauser2018mapping} can be localized in a disorder potential, each  with the localization quantum number $n=\text{1s, 2s, ...}$, cf. Fig. \ref{schema}(b). The center-of-mass localization approach  leading to new states below the free excitons   is consistent with the observed mid-bandgap states in the electron-hole picture in density functional theory studies on defects  \cite{salehi2016atomic,zhang2017defect,refaely2018defect}.

\begin{figure}[t!]
  \begin{center}
\includegraphics[width=\linewidth]{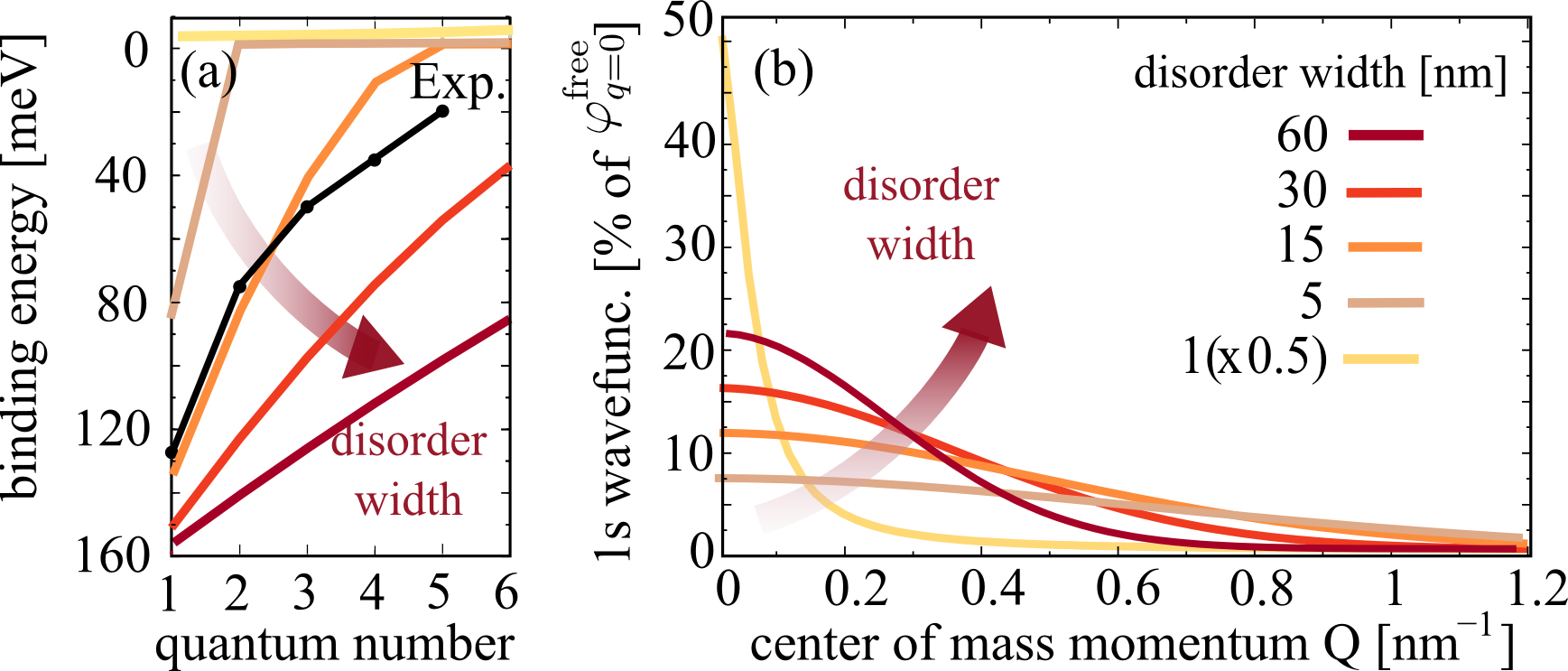} 
\end{center}
    \caption{\textbf{Characteristics of localized excitons.} Disorder-width dependent (a) binding energies and (b) 1s wavefunctions of localized excitons. The broader the disorder potential, the more localized states exist and the stronger is their binding energies. The calculated results are in good agreement with recent photoluminescence excitation measurements \cite{tonndorf2015single} (black line). }
   \label{bild2}
\end{figure}

To obtain access to the wavefunctions and eigenenergies of localized excitons, we solve the eigenvalue problem
\begin{equation}
 \left(\epsilon_{\mu} + \frac{\hbar^2 Q^{2}}{2M_{\mu}} \right)  \chi_{\bf Q}^{\mu n} 
 - \sum_{\bf Q'}  V_{\bf Q-Q'}^\mu \chi_{\bf {Q'} }^{\mu n}=\varepsilon_{\mu n } \chi_{\bf {Q}}^{\mu xn}
 \label{wannier}
\end{equation}
with the Fourier transformed disorder potential $V_{\bf Q-Q'}^\mu=V_0 \frac{\sigma ^2}{4\ln 2} \, \exp{\left(  \frac{-( { \bf Q-Q'})^2 \, \sigma^2}{16\ln 2}\right)} \varphi^{\mu*}_Q \varphi^\mu_{Q-Q'}$, including the formfactor $\varphi^{\mu*}_Q \varphi^\mu_{Q-Q'}$ stemming from the change into free exciton basis . Depending on the exact form of the disorder potential, we find bound or free solutions of this eigenvalue problem, corresponding to trapped or free excitons, respectively. 
For reasons of clarity, we fix the disorder potential to $V_0=120 \text{ meV}$ and vary the disorder width between 1 and 60 nm, cf. Fig. \ref{bild2}. The value of $V_0=120 \text{ meV}$ is chosen in agreement with recent photoluminescence excitation measurements \cite{tonndorf2015single}. For a very narrow potential in real space, excitons cannot be trapped due to their finite Bohr radius of approximately 1 nm \cite{berkelbach2013theory}. As soon as the potential is broad enough, trapping occurs and bound states appear (Fig. \ref{bild2}(a)). 
The broader the disorder potential in real space, the more bound eigenenergies fit into the potential. However, we also observe that the binding energies show a non-linear increase, i.e. when increasing from 30 to 60 nm the binding energy changes only slightly. If the potential is very wide in real space, it does not act as a strong local confinement anymore: excitons can be trapped inside but since the potential is so wide, excitons tend to behave more like free excitons, i.e. become broad in real space.
In momentum space, 
the wavefunctions become narrower and higher (Fig. \ref{bild2}(b)) resembling more and more the shape of free excitons. \\

\section{Photoluminescence of localized excitons}
The approach of disorder-induced center-of-mass localization allows us to write the equations of motion in a localized exciton basis. The advantage  is  that we can exploit the TMD Bloch equations for phonon-assisted photoluminescence derived in our previous work \cite{brem2019phonon}, transfer the PL equation in the localized exciton basis and obtain the PL intensity $ I^{b}_{\textbf{PL}}$ for localized states belonging to the bright exciton $X$:
\begin{equation} \label{eq:PLB} 
 I^{b}_{\text{PL}}(\omega)\propto \sum_{n} \dfrac{|M^{\text{b} n}|^2 \gamma^{\text{b} n} N^{\text{b} n} }{(\varepsilon_{\text{b} n}-\omega)^2+(\gamma^{\text{b} n}+\Gamma^{\text{b} n})^2}.
\end{equation}
Here, we use the index $b$ for the bright KK excitons within the light cone and $n$ for the quantum number of the localized state. 
The equation resembles the well-known Elliott formula \cite{hoyer2005many}. The position of  excitonic peaks is determined by the energy $\varepsilon_{bn}$, while the peak width  is given by radiative ($\gamma^{bn}$) and non-radiative dephasing ($\Gamma^{b,n}$)  due to phonons. The oscillator strength is determined by the optical matrix element, which reads in the new basis $|M^{\text{b} n}|^2 = \sum_Q | \chi_Q^{n}|^2  \delta_{Q,0}\sum_q  |\varphi_q |^2 M_q$ with localized and free exciton wave functions $\chi$ and $\varphi$, respectively, as well as the original momentum-dependent optical matrix element $M_q$ \cite{gunnar_prb}. The delta distribution assures that only bright excitons within the light cone with $Q=0$ are visible.
In the limit of no localized excitons, we find $\sum_Q | \chi_Q^{n}|^2  \delta_{Q,0} = 1$, which results in  the free-exciton photoluminescence formula \cite{brem2019phonon}. 

Moreover, our calculations show that the  wave functions of localized excitons are an order of magnitude weaker compared to  free excitons (Fig. \ref{bild2}(b)), and hence the oscillator strength of localized excitons  $|M^{\text{b} n}|^2$ becomes rather small. On the other side, the exciton occupation $N^{\text{b}n}$ of the energetically lower localized states is high at low temperatures. The occupation is determined by the capture rate of excitons into the considered disorder potential. The capture process has been treated microscopically  taking into account exciton-phonon scattering processes between free and localized excitonic states within the second-order Born-Markov approximation \cite{walls2007quantum} and applying the basis of orthogonal plane waves for localized states  \cite{herring1940new,schneider2003influence}. More details can be found in the supplementary material.

Interestingly, the reports in literature show that localized excitons are most pronounced in tungsten-based TMDs \cite{zhang2015experimental,cadiz2017excitonic}. The latter are characterized by the presence of momentum-dark K$\Lambda$ and KK' excitons  that are located below the  bright KK excitons \cite{malic2018dark,berghauser2018mapping}. Their role in presence of disorder has not been investigated yet. Depending on their energetic position and the ratio between capture and phonon-assisted intervalley scattering rates, the free KK excitons can either first be captured in a disorder potential or first scatter into a dark intervalley exciton state (K$\Lambda$, KK') and then be captured, cf. Fig. \ref{schema}(b). 
To account for these additional dark exciton scattering channels, we extend Eq.(\ref{eq:PLB}) including now phonon-assisted PL $I_{\text{PL}}^{d}$ from the states  $\nu= \text{KK, K}\Lambda, \text{KK}^\prime$ 
 \cite{brem2019phonon}:
\begin{equation} \label{eq:PLD} 
  I_{\text{PL}}^{d} (\omega)\propto \sum_{\substack{
\nu n \\
 \mu m \\
 Q,\alpha\pm}} 
 \Omega^{\nu n} (\omega)
 \dfrac{  |D^{\nu n {\mu m}}_{Q\alpha}|^2 \Gamma^{\mu m}  N^{\mu m} 
\eta_\alpha^{\pm}
}{(\varepsilon^{\mu m}\pm\Omega^\alpha_Q -\omega)^2 +(\Gamma^{\mu m})^2}. 
\end{equation}
Here, we have introduced the abbreviation $\Omega^{\nu  n}(\omega)=\frac{|M^{\nu  n}|^2 }{(\varepsilon_{\nu n}-\omega)^2+(\gamma^{\nu  n}+\Gamma^{\nu  n})^2}  $.
The position of excitonic resonances in the PL spectrum is now determined by the energy of the exciton $\varepsilon^{\mu m}$ and  the energy of the involved phonon $\pm\Omega^{\alpha}_{Q}$. The sign describes either the absorption (+) or emission (-) of phonons. We take into account all in-plane optical and acoustic phonon-modes. Moreover, the appearing phonon occupation $\eta_\alpha^{\pm} = \left(
\frac{1}{2} \mp \frac{1}{2} + n^{\text{phon}}_{\alpha}
\right)$ is assumed to correspond to the Bose equilibrium distribution according to a bath approximation \cite{axt1996influence}. Since dark states can not decay radiatively, the peak width is only determined by non-radiative dephasing processes $\Gamma^{\mu n}$. The oscillator strength of phonon-assisted peaks scales with the exciton-phonon scattering element  $|D^{\nu n {\mu}m}_{Q \alpha}|^2 = \sum_{Q'} \chi_{Q'+Q}^{\nu n*}  g^{\nu  \mu }_Q \chi^{\mu m}_{Q'}$, where $g^{\nu  \mu}_Q$ is the exciton-phonon coupling element  including free exciton wavefunctions \cite{selig2018dark,brem2019phonon}.

In this  study, we focus on processes involving the bright KK excitons as initial states, as those states are expected to be the most dominant in PL.
Note that in \eqref{eq:PLD}, the sum over $n$ describes the additive contribution of excited localized excitonic states. However,  our microscopic calculations show that the intraexcitonic scattering of localized excitonic states ($\mu n \rightarrow \mu n'$)  appears on a much faster timescale than the capture processes itself. Hence, it is sufficient to take into account only the ground 1s state for the calculation of the optical response. This  reduces the complexity to determine the exciton occupations $N^{\nu n}$ appearing in \eqref{eq:PLB} and \eqref{eq:PLD}. Taking into account all scattering processes on a microscopic footing, we calculate the exciton densities within the states $X,X_{\text{Loc}},X^D,X^{D}_{\text{Loc}}$, cf. Fig.\ref{schema}. We would like to emphasize that we calculate both the exciton decay via intervalley scattering $X\rightarrow X^{D}$ and the capture process $X^{(D)} \rightarrow X^{(D)}_{\text{Loc}}$ on a microscopic level driven by the exciton-phonon coupling strength $|D^{{\mu n} {\nu n'}}_{Q\alpha }|^2$, cf. the supplementary material for more details.

\begin{figure}[t!]
  \begin{center}
\includegraphics[width=\linewidth]{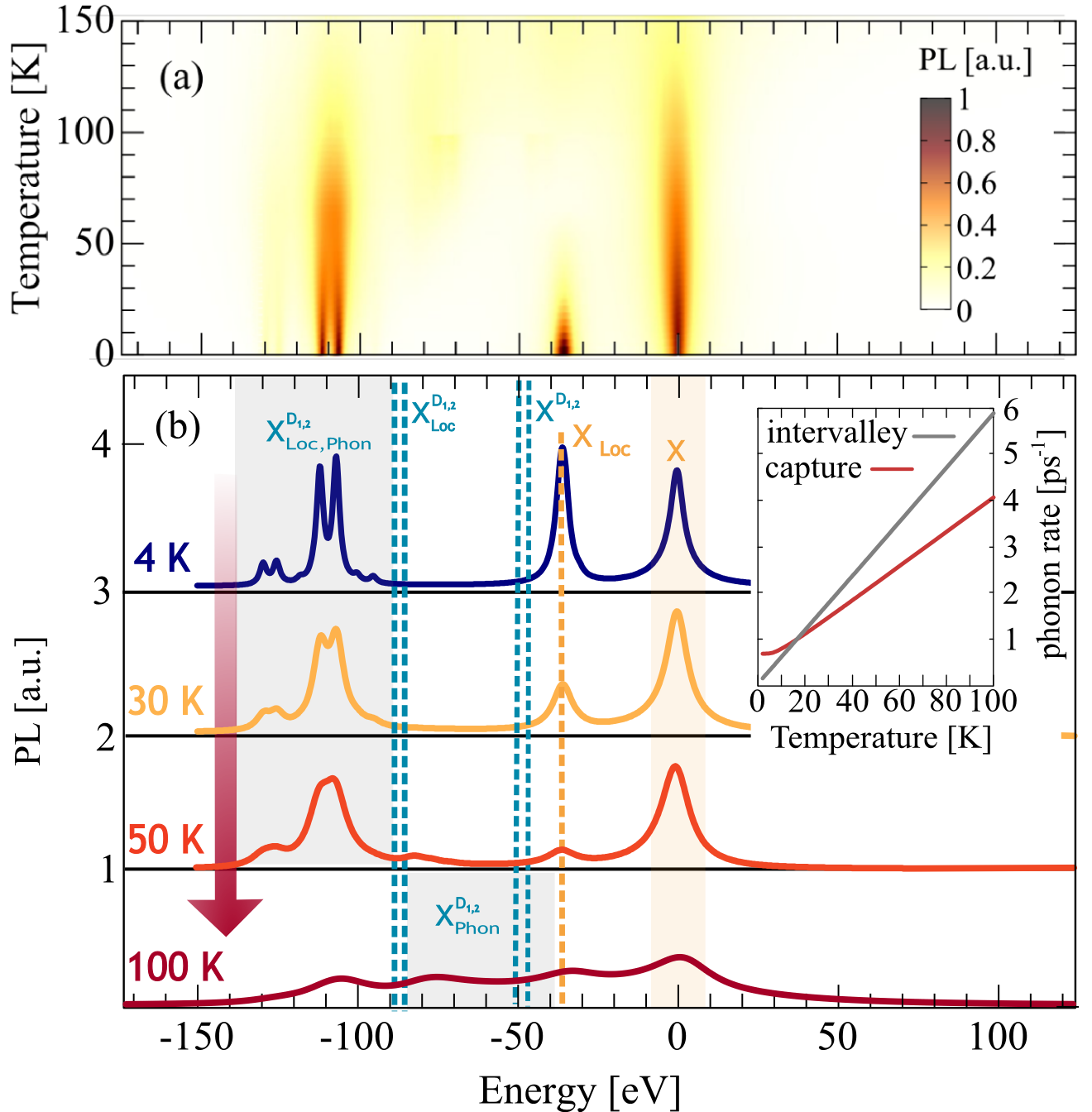} 
\end{center}
    \caption{\textbf{Temperature dependent photoluminescence.} (a) Surface plot of time integrated photoluminescence for different temperatures and  a fixed disorder potential with $\sigma=30 \text{ nm}$ and $V_0=40 \text{ meV}$. (b) 2D cuts from the surface plot at constant temperatures. Note that the spectra are shifted in energy so that the bright $X$ exciton is located at 0 meV. We observe a variety of new disorder-induced peaks below the bright exciton including $X_{\text{Loc}}$ and phonon-side bands of the momentum-dark localized excitons $X^D_{\text{Loc}}$. The inset shows phonon-induced capture (red) and intervalley (gray) scattering rates. For low temperatures (T$<$20 K) capture is dominant and hence the $X_{\text{Loc}}$ resonance is strongly pronounced.}
\label{bild3}
\end{figure}

Now we have all ingredients to investigate the photoluminescence of free and localized bright and dark excitonic states, i.e.  $  I_\text{PL} (\omega)= I_\text{PL}^{b} (\omega)+I_\text{PL}^{d} (\omega)$. Figure \ref{bild3} shows the time-integrated and temperature dependent PL spectra for the exemplary material tungsten diselenide ($\text{WSe}_2$) choosing  a fixed disorder width of $30 \text{ nm}$ and a disorder depth of $V_0=40 \text{ meV}$. These values are chosen by taking into account recent experimental measurements on nanopillars and deterministic local strain in TMD samples \cite{palacios2017large}. 
We observe a variety of peaks at low temperatures, which we can microscopically ascribe to: (i) the bright KK exciton $X$, (ii) the localized KK exciton $X_{\text{Loc}}$ (located about 40 meV below the $X$ resonance), (iii) a series of peaks around 100-130 meV below the $X$ resonance reflecting phonon-assisted PL from dark localized states $X^{D_{1,2}}_{\text{Loc}}$ ($D_{1,2}= \text{K}\Lambda, \text{KK}'$). In particular, we predict pronounced phonon side bands from the energetically lowest KK' exciton involving TO/LO phonons (around 130 meV) and TA/LA phonons (around 115 meV). Our findings support the hypothesis of recent experimental findings, predicting photoluminescence resonances stemming from dark localized excitons \cite{tripathi2018spontaneous,he2016phonon}.

We find that with the increasing temperature the linewidth of all peaks increases (Fig. \ref{bild3}(b)) due to the enhanced exciton-phonon scattering. For free excitons, this is intuitive and well established in literature \cite{selig2016excitonic}. However, for localized excitons the temperature behavior is still not well understood. Our calculations show (i) a temperature-independent contribution due to radiative decay  and (ii) an linear increase in the linewidth due to the enhanced exciton-phonon scattering. The radiative part is determined by the wavefunction overlap and we find $\gamma_0 \approx 0.5 - 1.0$ meV, which is in good agreement with the experimental obtained value of 0.9 meV \cite{he2016phonon}. The  increase with temperature stems from the temperature dependence of the phonon-driven capture rate, cf. the inset in Fig.\ref{bild3}. Even at 0 K we find a contribution to the capture rate due to phonon emission processes. This agrees well with the overall temperature behavior of localized excitons observed in experiments, revealing stable localized excitons up to 100 K \cite{he2016phonon,klein2019atomistic}.

Furthermore, we find that the localized bright exciton $X_{\text{Loc}}$ shows a fast intensity decrease with temperature. This  can be explained by the behavior of the capture rate (cf. the inset in Fig. \ref{bild3}(b)): For temperatures below 20 K the capture of excitons is faster than the intervalley scattering with phonons, and hence the intensity of   $X_{\text{Loc}}$  is high. The calculated capture rates are in good agreement with other theoretical studies \cite{ayari2019dynamics}. For temperatures above 20 K, the intervalley exciton-phonon scattering becomes faster, which means that excitons are more likely to scatter to the K$\Lambda$ and KK' states first, before they are captured in the respective localized states. This presents a crucial difference between bright and dark localized states, and their temperature behavior could be used to trace back their origin. Moreover, as  $X_{\text{Loc}}$ signature disappears with temperature, the probability to occupy the dark states $X^D$ increases and from about 50K we can see  traces of phonon side\cite{selig2018dark,brem2019phonon}bands from these actually dark states. We find additional peaks between 100  and 50 meV reflecting phonon emission and absorption processes from KK' and K$\Lambda$ excitons. To sum up, at low temperatures ($<$100 K) the PL is dominated by disorder-driven signatures, while at higher temperatures phonon-dominated signatures determine the optical response.

\begin{figure}[t!]
  \begin{center}
\includegraphics[width=\linewidth]{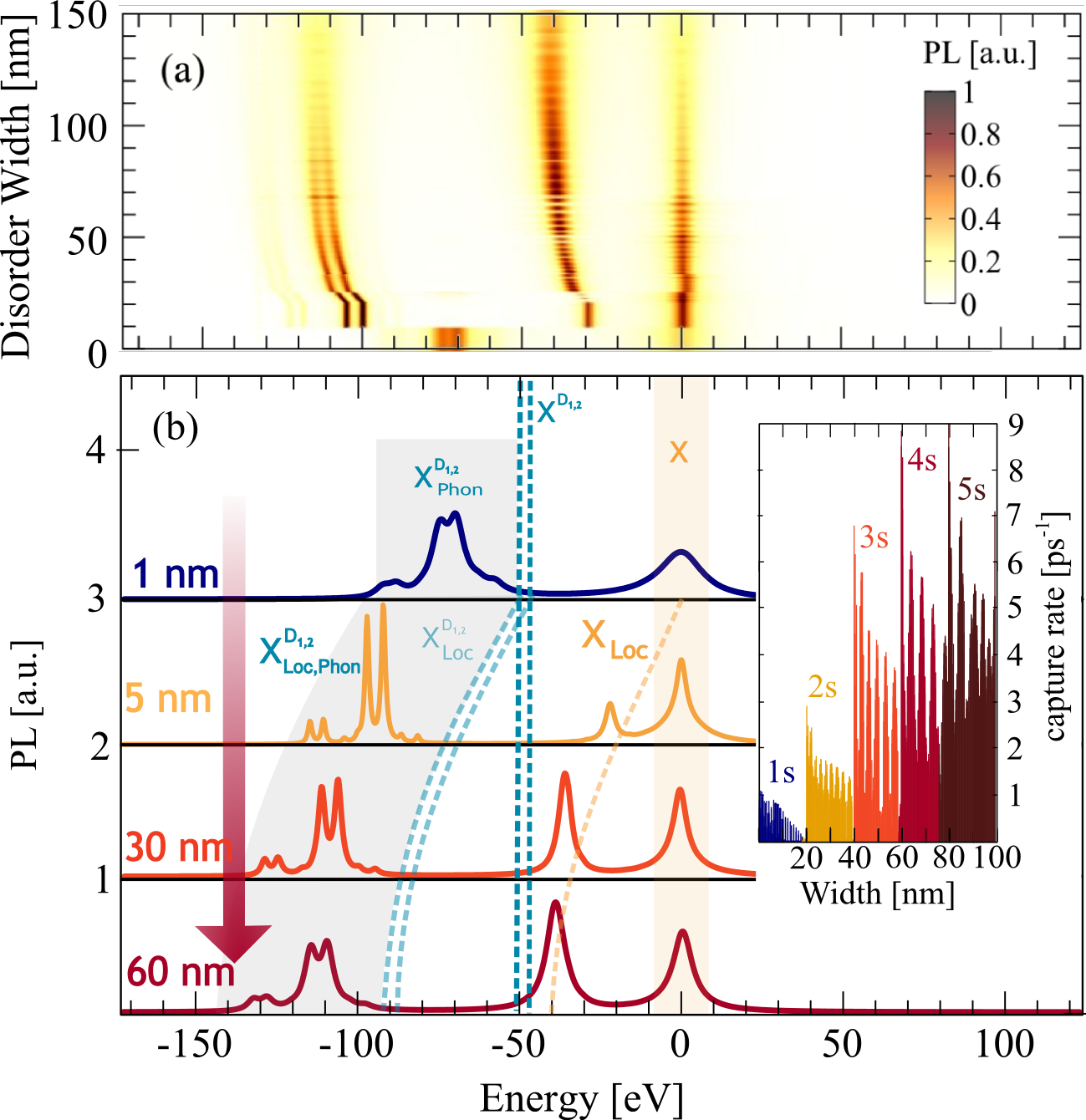} 
\end{center}
    \caption{\textbf{Disorder-width dependent photoluminescence.} (a) Time-integrated photoluminescence spectra at 10 K for varying disorder width. (b) 2D cuts of the surface plot at four fixed widths.
 If the disorder potential is too narrow (1 nm, blue line),  excitons can not be trapped and hence PL is dominated by the bright X exciton and phonon-assisted side bands of momentum-dark excitons $X^D$. For larger disorder widths, excitons become trapped giving rise to new peaks stemming from localized excitons. We predict clear signatures from both localized bright and dark excitons $X_{\text{Loc}}$ and $X_{\text{Loc}}^{D}$. For increasing disorder, we find that these peaks shift to lower energies due to deeper localization. Moreover,  while for a width of 5 nm phonon-assisted side bands from $X_{\text{Loc}}^{D}$ are predominant,  bright localized excitons $X_{\text{Loc}}$ gain intensity for larger disorder widths reflecting the increasing capture rate, cf. the inset.}
   \label{bild4}
\end{figure}

\section{Disorder-induced control of PL}
Having understood the origin of different resonances in low-temperature PL spectra in presence of disorder, the goal now is to investigate to what extent the spectra can be tuned by the characteristics of the disorder potential. We keep the temperature fixed at 4K, as this is the temperature with the most pronounced localized states, and vary the width of the potential. For a better understanding of the underlying processes, we keep the disorder depth constant ($V_0=const=40 \text{ meV}$). Changing the disorder depth would not change the qualitative behavior,  since the crucial quantity is the product of width $\sigma^2$  and depth $V_0$ appearing in the Fourier-transformed disorder potential, cf. Eq. (\ref{wannier}).
The strength of our approach is that we can describe different regimes including very narrow and deep disorder potentials and as well as disorders of width and height in the range of 100 nm and more.

As expected from the calculated eigenenergies in Fig. \ref{bild2}(a), we observe clear spectral shifts of localized excitons towards lower energies with increasing disorder width, cf. Fig. \ref{bild4}. However, at about 60 nm a saturation is reached and the position of the peaks remains constant. This behavior can be understood as follows: For very narrow disorder potentials (in the range of 1 nm and below), i.e. $\sigma < r_B$ with $r_B$ as  the exciton Bohr radius, excitons do not get trapped and hence we only observe  phonon side bands (50-100 meV below $X$) from the free dark excitons $X^{D_1,D_2}$ (Fig. \ref{bild4}(b) upper panel). Here, we are in the the phonon-dominated regime. As soon as the disorder potential is broad enough, excitons can be captured and localized bright excitons $X_{\text{Loc}}$ as well as phonon side bands of localized dark excitons $X^{D_1,D_2}_{\text{Loc}}$ appear (Fig. \ref{bild4}(b) lower panels). Here, we enter a disorder-dominated regime in the PL. When increasing the disorder width in real space, first excitons become stronger localized, however at some width a saturation is reached. This is due to the fact that with very wide potentials,  excitons are not tightly confined anymore and tend to behave like  free excitons again. They are characterised by a broad wavefunction in real space (and hence a narrow wavefunction in  momentum space, cf. Fig. \ref{bild2}(b)).

Beside the spectral shift discussed above, we also observe a clear disorder-induced increase of the resonance linewidth of localized excitons. This is due to narrower exciton wavefunctions (cf. Fig. \ref{bild2}(b)) and hence stronger radiative dephasing, as the latter directly scales with $ |\chi^{\mu n}|^{2}$. 
Interestingly, we also find that the intensity of localized dark and bright states clearly changes with the disorder width. This can be ascribed to the width dependence of capture processes, cf. the inset in Fig. \ref{bild4}(b). We find an overall enhanced capture efficiency at larger disorder widths due to enhanced overlap of the free and localized wavefunctions.  As we increase the disorder width, the localized wavefunctions become narrower and hence more similar to the free exciton wavefunction resulting in a larger  overlap.
 However, the capture rate is not only determined by the wavefunctions,  but it also crucially depends on the energetic position of the highest localized state. At low temperatures, free excitons can scatter only by emission of a phonon into the energetically lower localized state. Therefore, the energetic difference between free and localized state has to be in the range of the phonon energy ($\approx$ 20-30 meV). For larger disorder widths, the spectral distance between free and localized state increases  and hence phonon-driven  scattering becomes less probable. However, at some disorder width, an additional localized state emerges, which offers new scattering channels for phonons. Whenever this occurs, we observe a new peak in the capture rate (cf. different colors in the inset in Fig. \ref{bild4}(b)).

In conclusion, we have developed a microscopic approach to describe trapping processes of excitons in atomically thin 2D materials. We have calculated wavefunctions and energies of localized excitons depending on characteristics of the disorder potential. Using this knowledge, we have determined the optical fingerprint of the exemplary tungsten-based TMDs. We find a variety of pronounced peaks below the bright exciton stemming from both bright and momentum-dark localized states. The predicted signatures are strongly sensitive to temperature. We find a disorder-dominated regime at low temperatures and a free exciton phonon-dominated regime at higher temperatures.
 The gained insights can be extended to  a broader class of 2D materials and might help to design tailored trapping potentials.
\\

This project has received funding from the Swedish
Research Council (VR, project number 2018-00734) and the European Union's Horizon
2020 research and innovation programme under grant
agreement No 785219 (Graphene Flagship).

\bibliographystyle{apsrev4-1}
\bibliography{refererences_maja2}

\widetext
\clearpage
\begin{center}
\textbf{\large Optical fingerprint of bright and dark localized excitonic states in atomically thin 2D materials}\\
\vspace{.5cm}
\large{Maja Feierabend, Samuel Brem and Ermin Malic \\
\textit{Chalmers University of Technology, Department of 
Physics, 412 96 Gothenburg, Sweden}\\}
\vspace{.5cm}
 \textbf{\large {\textsc{-Supplementary Material - }}}\\
\end{center}
\setcounter{equation}{0}
\setcounter{figure}{0}
\setcounter{table}{0}
\setcounter{page}{1}
\makeatletter
\renewcommand{\theequation}{S\arabic{equation}}
\renewcommand{\thefigure}{S\arabic{figure}}
\renewcommand{\bibnumfmt}[1]{[#1]}
\renewcommand{\citenumfont}[1]{#1}

\section{Equations of motion in excitonic basis and Capture Rates}
The dynamics of exciton densities appearing in \eqref{eq:PLB} in the main text follow from the TMD Bloch equations \cite{feierabend2018molecule,selig2018dark} and can be written in its general form :
\begin{eqnarray}\label{Neq}
\dot{N}^{\mu }_\mathbf{Q}&=& 
\sum_{\nu, \bf{Q'}} \Gamma_{{\bf Q'}{\bf Q}}^{ \nu  \mu , \text{in}} |P^{\nu }_{{\bf Q'}} |^2 \delta_{{\bf Q', 0}}
-\Gamma^{\mu }_{\text{rad}} N^{\mu }_{{\bf Q}} \delta_{{\bf Q, 0}} 
+
 \sum_{\nu , \bf{Q'}} \left(
\Gamma_{{\bf Q'}{\bf Q}}^{ \nu  \mu , \text{in}} N^{\nu }_{{\bf Q'}} 
-\Gamma_{{\bf Q}{\bf Q'}}^{  \mu  \nu , \text{out}} N^{\mu }_{{\bf Q}} 
\right) 
\end{eqnarray}
The dephasing of the coherence $P^{\nu}_{{\bf Q'}}$ leads to the formation of incoherent excitons. The first contribution proportional to $\propto |P^2|$ is the driving term for the considered dynamics of excitons.
The incoherent excitons can decay radiatively with the rate $\Gamma^{\mu}_{\text{rad}}$ (second contribution in Eq. \ref{Neq}), as long as they are located within the light cone with $\textbf Q\approx 0$. This is valid for both free and localized KK excitons.
 Moreover, the incoherent excitons thermalize towards a thermal Bose distribution through exciton-phonon scattering (third contribution in Eq. \ref{Neq}). This is determined by  out-scattering rates $\Gamma^{\mu\nu, \text{out}}_{{\bf{QQ'}}}$ describing phonon-driven scattering from the state $(\mu, {\bf Q})$ to the state $(\nu, {\bf Q'})$ and in-scattering rates $\Gamma^{\nu\mu , \text{in}}_{{\bf{Q'Q}}}$ describing the reverse process. 
Transforming now in the localized exciton basis, we find for the dynamics of localized excitons:
\begin{eqnarray}\label{NeqL}
\dot{N}^{\mu m }&=& \sum_{\nu n}  \Gamma^{\nu n \mu m} | P^{\nu n}| ^2 - \Gamma_{\text{rad}}^{\mu } |\chi_{Q=0}^{\mu m}|^2 N^{\mu m} 
+
\sum_{\nu n}  \left( \Gamma^{\nu n \mu m} N^{\nu n} - \Gamma^{\mu m \nu n} N^{\mu m} \right)
\end{eqnarray}
with  exciton-phonon scattering rate \cite{selig2018dark} in localized exciton basis 
\begin{eqnarray}\label{scatti}
\Gamma^{\nu n \mu m}= \frac{2\pi}{\hbar} \sum_{\alpha Q'}| D^{\nu n \mu m}_{\alpha Q'} |^2 
\left(
\frac{1}{2} \mp \frac{1}{2} + n^{\text{phon}}_{\alpha Q'}
\right)
\delta(\varepsilon^{\mu m} -\varepsilon^{\nu n} \pm \Omega^\alpha_{Q'}) \quad.
\end{eqnarray}
where $D^{\nu n \mu m}_{\alpha Q'}$ corresponds to the exciton-phonon matrix elements ,$n^{\text{phon}}_{\alpha Q}$ describes the phonon occupation, $ \Omega^\alpha_Q$ denoting the energy of the involved phonon, and $\varepsilon^{\mu m} $corresponding to the exciton energy  of the involved states.
The delta distribution in \eqref{scatti} assures energy conservation between initial and final exciton state under emission/absorption of phonons.

The appearing exciton-phonon matrix elements read 
\begin{equation}\label{raten}
D^{\nu n \mu m}_{\alpha Q}=\sum_{qQ'} \chi_{Q}^{\nu n*}  
  \varphi_q^{\nu*} g^{cc}_{Q'\alpha} \varphi_{q+\beta Q'}^{\mu} 
 \chi^{\mu m}_{Q+Q'}
\end{equation}
including both free $\varphi$ and localized $\chi$ wavefunctions and the electron-phonon coupling elements $g^{cc}_{Q\alpha}$ \cite{brem2019intrinsic} . Depending on initial and final state we can distinguish three processes: (i) $\text{Free} \rightarrow \text {Free}$ described by $\Gamma_{{\bf Q'}{\bf Q}}^{ \nu \mu  , \text{phon}}$, (ii) $\text{Loc} \rightarrow \text {Loc}$ described by $\Gamma^{ \nu n \mu m , \text{loc}}$, and (iii) $
\text{Free} \rightarrow \text {Loc} \, (\text{Loc} \rightarrow \text {Free})$ described by $ \Gamma^{ \nu F \mu m , \text{capt}} (\Gamma^{ \mu m \nu F, \text{esc}})$.
Scattering between free exciton states  includes both intra- and inter-valley scattering. 
Scattering between localized exciton states is restricted to processes within the same valley, since intervalley scattering would involve at least two phonons. Phonon-driven scattering from a free to a localized state corresponds to a capture or an escape process. 

Note that our system has to be in an orthogonal basis, i.e. it has to yield $\Phi_{qQ}=\chi_Q\phi_q$ where $\Phi_{qQ}$ are eigen vectors of an orthogonal system. To describe the continuum of states, we use plane waves and  describe the free eigenfunctions by orthogonalized plane waves \cite{herring1940new,schneider2003influence,malic2006theory}:
\begin{equation}
| \phi^{OPW}_Q \rangle =\frac{1}{N_Q^{}}
\left(
| \phi_Q^{PW} \rangle - 
\sum_{\mu m}
\langle
\chi^{\mu m}_Q | \phi_Q^{PW} \rangle | \chi^{\mu m}_Q \rangle
\right)
\end{equation}
with the normalization factor $N_Q^{} = \sqrt{1- \sum_{\mu m}|\langle \chi^{\mu m}_Q | \phi_Q^{PW} \rangle| ^2}$ and the plane waves $\phi^{PW}_Q$ which can be described in momentum space by a delta function around the $Q_F$, ie.  $\phi^{PW}_Q \approx \delta_{Q,Q_F}$. 
Using this approach,  we are able to calculate all scattering and capture rates.  We find that the capture of excitons is most likely to happen in the energetically closest localized state. Capture in energetically lower lying states appears on a much slower timescale and is hence negligible. The reason for that is the energy conservation in \eqref{raten}. 
Furthermore, the relaxation dynamics within the localized states happen on a much faster time scale than capture processes, i.e. excitons decay almost immediately from any $n$s state to the lowest 1s state. Hence, it is most important to take into account the localized 1s states for the calculation of the optical response. 

Assuming a Boltzman distribution for the free states $N^{\mu F}_{Q_F}\approx N^{\mu F}_{0} e^{-\beta (E^{\mu F}_{Q_F}-E^{\mu F}_{0})} $ yields to $\sum_{Q_F} N^{\mu F}_{Q_F} = N^{\mu F}_{0} \kappa_\mu$ with $\kappa_\mu=\frac{M_\mu}{2\pi\hbar^2\beta}$ which enables us to evaluate \eqref{Neq} and \eqref{NeqL}. Taking all intra- and intervalley exciton-phonon scattering as well as capture and escape processes into account, we  find for the dynamics of bright and momentum-dark exciton $\mu$ = K$\Lambda$,KK'), both free $F$ and localized $L$ states :
\begin{eqnarray}\label{Neq2}
 \dot{N}^{\mu F}_{0} &=& 
\frac{1}{\kappa_\mu} 
\left[
\Gamma^{\mu F,\text{form}} 
-\left(\Gamma^{\mu F,\text{rad}}  +\Gamma^{\mu F \mu L \text{capt}}  \right)
{N}^{\mu F}_{0}
+ 
\Gamma^{\mu L \mu F \text{esc}}
{N}^{\mu L}
+
\sum_{\nu F}
\left(
\Gamma^{\mu F \nu F \text{in}} {N}^{\nu F}_{0}
-
\Gamma^{\nu F \mu F \text{out}} {N}^{\mu F}_{0}
\right)
\right] \\
 \dot{N}^{\mu L} &=& 
\Gamma^{\mu L,\text{form}} 
-\left(\Gamma^{\mu L,\text{rad}}  +\Gamma^{\mu L \mu F \text{esc}}  \right)
{N}^{\mu L}
+ 
\Gamma^{\mu F \mu L \text{capt}}
{N}^{\mu F}_{0} 
\end{eqnarray}
where we have introduced the radiative dephasing $\Gamma^{\mu F(L),\text{rad}}$ for the free (localized) state within the light cone, ie. $\mu$=KK, a term $\Gamma^{\mu F(L),\text{form}} = \Gamma^{\mu F(L) \mu L} |P^{\mu L}|^2 + \sum_{\nu Q_F}  \Gamma^{\mu F(L) \nu Q_F} |P^{\nu F}|^2$ which corresponds to the driving term due to decay of coherent excitons,   $\Gamma^{\mu F \mu L \text{capt}} =\sum_{Q_F} \Gamma^{\mu F \mu L}_{Q_F}  e^{-\beta E^{\mu F}_{Q_F}} $ as capture and
  $\Gamma^{\mu L \mu F \text{esc}} =\sum_{Q_F} \Gamma^{\mu L \mu F}_{Q_F}  $ as escape rates, and finally for in and outscattering with phonons between free excitons $\Gamma^{\mu F \nu F \text{in}} =\sum_{Q_F Q'_F }  \Gamma^{\nu F \mu F}_{Q'_F Q_F} e^{-\beta E^{\nu F}_{Q'_F}}$ and $\Gamma^{\mu F \nu F \text{out}} =\sum_{Q_F Q'_F }  \Gamma^{\mu F \nu F}_{Q_F Q'_F} e^{-\beta E^{\mu F}_{Q_F}}$.\\
The radiative decay of both free and localized excitons is calculated by exploiting the corresponding wave functions obtained through the Wannier equation \cite{selig2016excitonic}. All appearing exciton-phonon scattering and capture/escape rates have been calculated on microscopic footing within the second-order Born-Markov approximation \cite{walls2007quantum}  and exploiting the orthogonalized plane wave approach  \cite{herring1940new,schneider2003influence,malic2006theory} as discussed above. Solving \eqref{Neq2} provides access to  time dependent exciton occupations in different  exciton states entering the photoluminescence formula \eqref{eq:PLB} in the main manuscript.

\end{document}